# MACHINE INTELLIGENCE-DRIVEN CLASSIFICATION OF CANCER PATIENTS-DERIVED EXTRACELLULAR VESICLES USING FLUORESCENCE CORRELATION SPECTROSCOPY: RESULTS FROM A PILOT STUDY.

[1]Abicumaran Uthamacumaran, [2,3]Mohamed Abdouh, [4]Kinshuk Sengupta, [5]Zu-hua Gao, [6]Stefano Forte, [2]Thupten Tsering, [2,7,8]Julia V Burnier and [9,10,11]*Goffredo Arena

[1]Concordia University, Department of Physics (Alumni), Montreal, QC, Canada

[2] Cancer Research Program, Research Institute of the McGill University Health Centre, 1001 Decarie Boulevard, Montreal, Quebec, Canada, H4A 3J1

[3] The Henry C. Witelson Ocular Pathology Laboratory, McGill University, Montreal, QC, Canada

[4] Microsoft Corporation, New Delhi, India

[5] Department of Pathology and Laboratory Medicine, University of British Columbia, G105-2211 Wesbrook Mall, Vancouver BC, Canada V6R 2B5

[6] iOM Ricera, via Penninazzo 11, 95029 Viagrande, Italy

[7]Gerald Bronfman Department of Oncology, McGill University, Montreal, QC, Canada

[8]Experimental Pathology Unit, Department of Pathology, McGill University, Montreal, Qc, Canada

[9] Istituto Mediterraneo di Oncologia, Viagrande, Italy

[10] Department of Surgery, McGill University, St. Mary Hospital, 3830 Lacombe Avenue, Montreal, Quebec, Canada, H3T 1M5

[11] Fondazione Gemelli-Giglio, Contrada Pollastra, Cefalu', Italy

*Correspondence:* Goffredo Arena* (Primary Supervisor):

goffredoarena@gmail.com

**ABSTRACT.**

**Background:** Patient-derived extracellular vesicles (EVs) that contains a complex biological cargo is a valuable source of liquid biopsy diagnostics to aid in early detection, cancer screening, and precision nanotherapeutics. In this study, we predicted that coupling cancer patient blood-derived EVs to time-resolved spectroscopy and artificial intelligence (AI) could provide a robust cancer screening and follow-up tools**.**

**Methods:** In our pilot study, fluorescence correlation spectroscopy (FCS) measurements were performed on 24 blood samples-derived EVs. Blood samples were obtained from 15 cancer patients (presenting 5 different types of cancers), and 9 healthy controls (including patients with benign lesions). EVs samples were labeled with PKH67 dye. The obtained FCS autocorrelation spectra were processed into power spectra using the Fast-Fourier Transform algorithm. The processed power spectra were subjected to various machine learning algorithms to distinguish cancer spectra from healthy control spectra.

**Results and Applications:** The performance of AdaBoost Random Forest (RF) classifier, support vector machine, and multilayer perceptron, were tested on selected frequencies in the N=118 power spectra. The RF classifier exhibited a 90% classification accuracy and high sensitivity and specificity in distinguishing the FCS power spectra of cancer patients from those of healthy controls. Further, an image convolutional neural network (CNN), ResNet network, and a quantum CNN were assessed on the power spectral images as additional validation tools. All image-based CNNs exhibited a nearly equal classification performance with an accuracy of roughly 82% and reasonably high sensitivity and specificity scores.

**Conclusion:** Our pilot study demonstrates that AI-algorithms coupled to time-resolved FCS power spectra can accurately and differentially classify the complex patient-derived EVs from different cancer samples of distinct tissue subtypes. As such, our findings hold promise in the diagnostic and prognostic screening in clinical medicine.

**INTRODUCTION.**

Cancers globally remain amidst the leading-cause of disease-related mortality. Conventional therapies may be successful for certain subtypes of the disease, while others are complex adaptive systems progressing to clinically aggressive stages causing a paramount disease burden. Further, the long-term health complications and side-effects, successfully treated patients must live with, must be emphasized. Within this pilot study, in efforts to advance precision oncology and patient-centered clinical medicine, we explored the application of artificial intelligence (AI) in tackling one of the greatest challenges in preventive and diagnostic medicine*: early cancer detection* and *prognostic screening*. Cancer biomarker discovery was pioneered by Gold and Freedman (1965) with their recognition of the first tumor marker, Carcinoembryonic Antigen (CEA), which remains to date the most used clinically-relevant, blood-based cancer screening and diagnostic in patient-care. Their co-discovery of the tumor-specific antigen gave birth to the field of precision immuno-oncology. Since then, significant progress has been made in the art of diagnostic medicine with the emergence of liquid-biopsies and longitudinal blood monitoring. Liquid biopsies are enriched with a complex variety of clinically-relevant information which can be exploited for robust biomarker discovery in cancer screening. Some of the rich sources of these markers include differential methylome signatures of cell-free circulating tumor DNA, cell-free RNA/microRNAs, circulating tumor cells (CTCs) (including quiescent/dormant cells), immune cells (and their population densities), immune cells-secreted signals and cytokines, and *extracellular vesicles* (EVs) (Alix-Panabières and Pantel, 2013; Han et al., 2017; Bronkhorst et al., 2019; Zhou et al., 2020; Sui et al., 2021). Among all of them, EVs, are emerging as a promising clinical candidate for robust, financially cheap, quick, and non-invasive liquid-biopsy characterization of cancer dynamics, clinical screening, disease progression monitoring, and patient-therapy management (Zhao et al., 2019; Zhou et al., 2020).

Early detection of cancer presents an interdisciplinary complex problem in diagnostic medicine. The longitudinal analysis of patient-derived tumor biopsy sequencing and molecular cytogenetics may be inaccessible to patients, due to their invasiveness and financial barriers. Further, there are limited antigen/biomarker tests and clinically-relevant blood-immune monitoring methods for complex adaptive cancers, such as aggressive brain tumors. Extracting CTCs or dormant cancer cells from disseminated tumors may be limited to only certain cancers at later stages of tumor progression. Then, the question arises: How do we sensitively detect cancers within patients at their early stages? How can we non-invasively perform longitudinal monitoring of therapy response in cancer patients? Medical physics applications such as CT/MRI-based imaging modalities, or the more painful lumbar punctures, are often limited to the detection of lesions with the presence of a minimal detectable size, the detection/profiling of later stages of disease progression, can be painfully invasive to patients, and present limitations. To reconcile the complex problem of early-stage cancer detection and screening, herein, we exploit complex systems physics and machine intelligence-driven pattern analysis in characterizing the time-resolved spectroscopic signals from patient sera-derived EVs.

*Complex systems theory* is the interdisciplinary study of quantifying the self-organized patterns and collective (emergent) behaviors in many-body nonlinear systems (i.e., complex systems) and processes (i.e., dynamical systems), by merging tools from artificial intelligence (AI), statistical physics, information theory, and nonlinear dynamics. EVs dynamics are *complex systems.* The field of EVs is rapidly evolving and different categories are now being recognized including exosomes, microvesicles, ectosomes, apoptotic bodies, etc. EVs are nanoscopic lipid-bound entities found in different bioliquids such as blood sera. Notably, they transmit intercellular information and regulate many physiological and pathological processes, such as controlling cancer cellular cybernetics. Rose Johnstone initiated the field of EVs physiology when she first characterized them through electron microscopy (EM) imaging of reticulocytes

(Pan et al., 1983; 1985; Johnstone, 2005). Our study will mainly seclude to the isolation of EVs from patient blood samples. Cells-secreted EVs are one of the primary cybernetic control systems mediating intercellular communication in physiological conditions. In vivo patient-EVs dynamics exhibit many complex adaptive features, including but not limited to the horizontal transfer of malignant traits, phenotypic reprogramming of distant tissue microenvironments into pre-metastatic niches, transcriptional and metabolic rewiring of cellular states, intracellular cargo transport, immune system control, regulating the phenotypic plasticity of cancer (stem) cells, conferring phenotypic heterogeneity in tumor microenvironments (TMEs), immunomodulation of tumor ecosystems, and promoting therapy-resistance (Abdouh et al., 2014; 2016; 2020; Arena et al., 2017; Zhou et al., 2017; Szatenek et al., 2017).EVs are also emerging as cell fate reprogramming nanotechnologies in precision nanomedicine. For instance, the EVs derived from cancer stem cells can form complex cell-cell communication networks which promote and dynamically remodel an immunosuppressive TME, and thereby confer therapy resistance in tumor ecosystems (Su et al., 2021). Patient blood-derived EVs provide a rich repertoire of complex information dynamics, due to the heterogeneity emerging from their multicellular origins, and their adaptive signals in response to their environmental perturbations. EVs are also emerging as patient-compatible, personalized nanotherapeutics and drug delivery vehicles (Fu et al., 2020). Further, it remains questioned whether in vitro reconstitutions of these complex systems may exhibit collective dynamics and emergent behavioral patterns due to their aggregate interactions (Uthamacumaran et al., 2022).

The application of AI is increasingly becoming prominent for pattern discovery in applications of precision medicine, ranging from automated multimodal drug discovery to blood/sera screening for complex disease markers. Precision medicine is now shifting towards the use of artificial intelligence, and in specific, statistical machine learning (ML) algorithms-driven pattern discovery in disease monitoring/screening. Statistical ML algorithms, including Deep Learning artificial neural networks, have been validated as robust tools for classification tasks/problems. There exists many examples of such types of works using AI and ML algorithms in liquid-biopsy based cancer biomarker discovery (Park et al., 2017; Shin et al., 2018; 2020; Uthamacumaran et al., 2022). In specific to AI applications to EVs profiling, a recent study demonstrated the merging of ResNet, a residual neural network-Deep Learning algorithm, and Surface-enhanced Raman spectroscopic characterization of liquid-biopsy derived EVs could yield >90% sensitivity and accuracy in cancer detection (Shin et al., 2020). These findings strongly suggest the pairing of liquid-biopsy derived cancer EVs with AI may pave a sensitive early-stage and prognostic detection of cancers in clinical medicine (Shin et al., 2020).

In extension to these findings, in our previous study we discovered that simple ML algorithms such as Random Forest (RF) classifiers and decision trees show high statistical accuracy in distinguishing the complex cancer patients-derived EVs Raman and FT-IR vibrational spectra from those of healthy patients (Uthamacumaran et al., 2022). Our study remains the first of such pilot studies to demonstrate the applicability of RF classifier, and similar ML algorithms, on patient sera-derived EVs' vibrational spectra (Uthamacumaran et al., 2022). While advanced Raman techniques such as SERS and Raman imaging, and an increased patient size with a diverse cancer subtypes/stages are required to further advance the clinical relevance of our findings, a fundamental limitation of such spectroscopic methods remains the lack of *time-series analysis* of the EVs temporal behaviors and features. In general, time-resolved spectroscopic techniques are under-investigated in the quantitative analysis of disease-driven complex systems, such as patient-derived EVs dynamics. Fluorescence Correlation Spectroscopy (FCS) is one such time-resolved technique in which we measure temporal fluctuations in fluorescently-labelled particles or chemical agents within a system to quantify its behavioral dynamics (Rigler and Elson, 2001). In this study, we demonstrate the first-time applicability of FCS in distinguishing cancer patient derived EVs from healthy patients. Within simple chemical systems, we assume the fluctuations to follow Brownian motion although complex systems can exhibit collective (aggregate), emergent behaviors (Uthamacumaran et al., 2022). Traditionally, the technique is used to quantify chemical characteristics of the system such as the diffusion coefficients, chemical kinetic rate constant, and molecular concentrations. Further, FCS allows the monitoring of ligand-macromolecule interactions with live-cell imaging at a single-molecule detection sensitivity (Thompson, 2002). As such, FCS provides a light-matter interaction interface to quantify complex systems dynamics, such as the chemical flow patterns of diffusive, molecular systems. A schematic of a generic FCS apparatus is shown in **Figure 1**. Herein, we exploit this technique to quantify temporal features in nanoscopic complex systems such as patient-derived EVs systems.

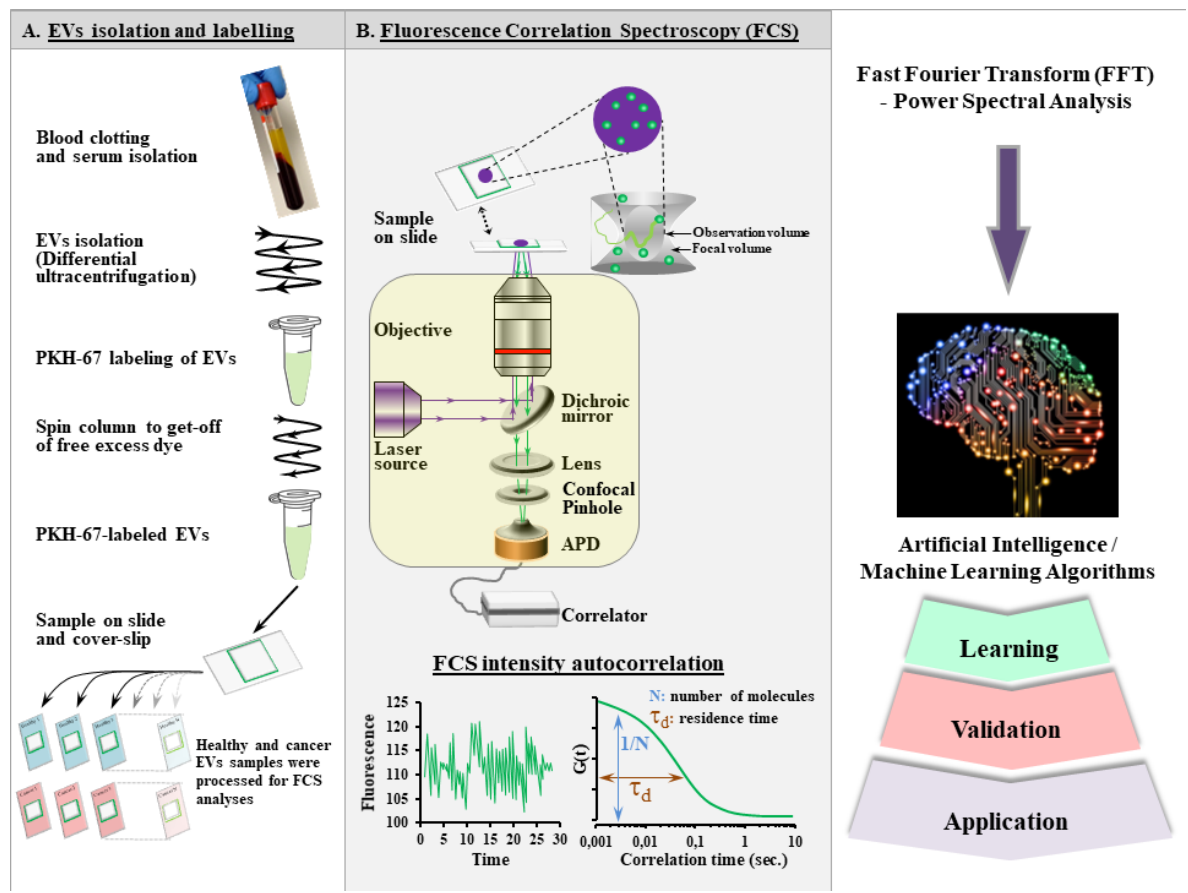

**FIGURE 1. WORKFLOW.** A schematic of the steps taken for the spectroscopic characterization of patient-derived EVs and pattern detection via Artificial Intelligence (AI). A) EVs isolation and membrane-fluorescent labelling with PKH67 for healthy controls and cancer patient-derived samples. B) FCS measurements were performed on the samples and the autocorrelation plots (vs. correlation time) were obtained from their fluorescence intensity fluctuations. C) The autocorrelation plots were subjected to the Fast-Fourier Transform (FFT) algorithm to obtain their power spectra. The power spectra exhibit finer spectral features which allowed optimal machine-driven classification. The power spectra were subjected to classification by various machine algorithms including machine learning (ML) classifiers, spectral image-based convolutional neural networks (Image CNN and ResNet), and an image-based quantum neural network (QNN). Statistical

measures were used as validation tools of the ML algorithms' performance and support applicability in clinical medicine.

At equilibrium, the fluorescent molecules undergo diffusive/flow processes within an illuminated opening/cavity (i.e., the focal volume), under the FCS microscope, giving rise to fluorescence intensity fluctuations over time. An autocorrelation function $G(\tau)$ is obtained as a function of the fluorescence decay time $\tau$, to quantify the average duration of the fluctuations. The autocorrelation function is given by:

$$G(\tau) = \frac{\delta F(t)\delta F(t+\tau)}{<F>^2}$$

Wherein $\delta F(t)$ denotes the fluctuations in the measured fluorescence F from the average fluorescence $<F>$ (Rigler and Elson, 1974; Thompson, 2002). The emitted intensity fluctuations are detected by the excited laser beam from the FCS apparatus, wherein the intensity is proportional to the number of fluorescently-labelled EVs molecules in the confocal volume (illuminated region). The flow dynamics and interactions of the EVs system, via diffusion, reaction, or other collective dynamics, causes the fluctuations to emerge (Elson and Magde, 1974). While traditional approaches to FCS analysis relied on extracting chemical and physical parameters from the autocorrelation function, we exploit herein complex systems tools, namely, FFT-power spectral analysis, multifractal analysis, and AI as complex feature extraction and classification approaches in the characterization of these time-resolved spectra. There remain a few studies which have used FCS to quantify EVs dynamics in healthy cellular systems. However, our study remains to date the first FCS application in cancer patients-derived EVs dynamics and demonstration of its clinical relevance to personalized nanomedicine.

FCS has been shown as a promising tool to quantify and visualize the EVs dynamics at the single-vesicle level of healthy cellular systems to elucidate cell to cell communication networks (Corso et al., 2019). FCS techniques in combination with other molecular translation techniques have been implemented in the profiling of EVs surface proteins in relation to their diffusion times of antibody-vesicle interactions (Fu et al., 2020). Wyss et al. (2014) used ultrafiltration and size-exclusion chromatography, as purification techniques to isolate EVs secreted by mammalian cells and used fluorescence fluctuation analysis by FCS to investigate their biophysical properties, such as diffusion times, in relation to EVs size distribution. However, there may be more optimal techniques such as nanoparticle tracking analysis (NTA) or dynamic light scattering (DLS) better suited for such size-exclusion analyses (Szatanek et al., 2017). We predicted that the temporal fluctuations of patient-derived EVs within the FCS confocal volume may provide insights into their temporal behaviors and collective dynamics, which remain presently unreported in disease systems. Our pilot study demonstrates for the first time that FCS fluctuations could provide clinically-meaningful insights into EVs dynamics and has the potential to accurately detect cancer EVs and be used in liquid biopsies.

**METHODS**

**Blood collection and serum preparation:** Patients for the current study were recruited form the department of General Surgery at the Royal Victoria Hospital and St-Mary's Hospital (Montreal, Canada) and underwent a written and informed consent for blood collection in accordance with protocols approved by the Ethics Committee of the McGill University Health Centre (MP-37-2018-3916 and SDR-10-057). Blood samples were collected from both healthy individuals and patients who presented to our clinic for a follow-up or those that underwent resection of primary cancer (**Table 1**). Blood samples (2 to 5 ml) were collected from a peripheral vein in vacutainer tubes (Becton Dickinson) containing clot-activation additive and a barrier gel to isolate serum. Blood samples were incubated for 60 min at room temperature to allow clotting and were subsequently centrifuged at 1500 g for 15 min. The serum was collected, and a second centrifugation was performed on the serum at 2000 g for 10 min, to clear it from any contaminating cells. Serum samples were aliquoted and stored at −80°C until further use.

**Table 1.** The profiles of blood donors (n=24).

| Patients ID | Case description | Age (years) | Sex |
|---|---|---|---|
| C1 | Healthy Control | 58 | Female |
| C2 | Healthy Control | 62 | Male |
| C3 | Healthy Control | 57 | Male |
| C4 | Healthy Control | 48 | Male |
| C5 | Healthy Control | 54 | Female |
| C6 | Healthy Control | 62 | Female |
| 244 | CYST | 40 | Male |
| 417 | Benign | 70 | Male |
| 101115 | Liver mass (benign) | 65 | Female |
| 269 | Cholangiocarcinoma | 77 | Male |
| 322 | Cholangiocarcinoma | 62 | Male |
| 332 | Cholangiocarcinoma | 73 | Female |
| 341 | Cholangiocarcinoma | 77 | Female |
| 497 | Cholangiocarcinoma | 58 | Female |
| 306 | Hepatocellular carcinoma | 76 | Male |
| 420 | Hepatocellular carcinoma | 68 | Male |
| 515.1 | Hepatocellular carcinoma | 59 | Female |
| 300 | Gallbladder cancer | 49 | Male |
| 498 | Ovary cancer with liver metastasis | 61 | Female |
| 298 | Colorectal cancer with liver metastasis | 74 | Male |
| 307 | Colorectal cancer with liver metastasis | 44 | Male |
| 335 | Colorectal cancer with liver metastasis | 75 | Female |
| 354 | Colorectal cancer with liver metastasis | 59 | Male |
| 431 | Colorectal cancer with liver metastasis | 64 | Female |

**Extracellular vesicles (EVs) isolation from serum samples**
Serum samples were diluted in phosphate-buffered saline (PBS) at 1/10 dilution and were subjected to a series of sequential differential centrifugation steps. Samples were centrifuged at 500 g for 10 min to remove any contaminating cells, followed by centrifugation at 2000 g for 20 min to remove cell debris. Supernatants were passed through a 0.2 μm syringe filter (Corning), transferred to 26.3 ml polycarbonate tubes (# 355618; Beckman Coulter), and centrifuged at 16,500 g for 20 min at 4°C to remove apoptotic bodies and cell debris. Supernatants were transferred to new 26.3 ml polycarbonate tubes and ultracentrifuged at 120,000 g (40,000 rpm) for 70 min at 4°C using a 70 Ti rotor in Optima XE ultracentrifuge machine (Beckman Coulter). The crude EVs pellets were washed with PBS at 120,000 g for 70 min at 4ºC, resuspended in 500 µl PBS, and stored in -80°C until further use.

**Extracellular vesicles (EVs) labeling:** Isolated EVs were labeled with PKH67 green fluorescent probe according to the manufacturer's instructions (Sigma). Briefly, EVs were resuspended in Diluent C and mixed with equal volume of the stain solution (4 ul PKH 67 in 1 ml Diluent C) for 5 min. The reaction was stopped by adding 2 ml of 2% bovine serum albumin or fetal bovine serum. Samples were passed through Exosome Spin Columns (MW 3000) (Thermo Fisher Scientific) to purify labeled EVs from unbound PKH67 dye. Samples were then centrifuged at 120,000 g for 70 min at 4°C. Labeled EVs pellets were resuspended in PBS for subsequent fluorescence correlation spectroscopy (FCS) analyses. For machine calibration, 2 controls were run in parallel: aliquots of PBS and diluted samples of PKH67 ($10^{-8}$M).

**Fluorescence Correlation Spectroscopy (FCS):** The FCS system from McGill University's ABIF (Advanced Bioimaging Facility) was used for our experiment. Fluorescence correlation spectroscopy measurements were performed at room temperature on a commercial Zeiss LSM 780 laser scanning confocal microscope with an inverted AxioObserver Z.1 stand and operated with Zen 2012 SP5 FP3 software including an FCS module (Zeiss) (**Figure 1**). We used a continuous wave 25 mW 488 nm argon laser and a 40X C-APOCHROMAT NA 1.2 W Korr UV-VIS-IR water immersion objective, with the correction collar adjusted for 0.17 mm cover glass at 23°C. Before each measurement session, a blank measurement was made with PBS for calibration. Samples were diluted in PBS in a Mattek 35 mm petri dish with a 14 mm microwell and a No 1.5 cover glass, and measurements were performed by focusing roughly 5 µm above the surface of the cover glass in the centre of the field of view. Laser intensity was controlled with an acousto-optic tunable filter set to 2% transmittance, the pinhole was set to 34 µm (as software recommended for 1 airy unit for this emission range and objective choice) and a 488 nm main beam splitter was used to separate excitation light from emission light. Raw photon counts were measured on a spectral detector with a range of 499-691 nm, and autocorrelation was calculated on the fly. Count rate binning was 1 ms, correlator binning time was 0.2 µs, and acquisition time was 30 s per run. Binned counts and calculated autocorrelation values were exported as ConfoCor3 fcs files and processed offline. Note: Use the R-language *ImportFCS* code to obtain the autocorrelation spectra from raw FCS intensity counts (See GitHub link for Data and Codes Instructions).

**Power Spectra:** The FCS autocorrelation spectra obtained from the ImportFCS code discussed above were further processed using the fast-Fourier transform (FFT) using OriginPro v 8.5. The Autocorrelation spectra data tables were inputted into the OriginPro software, and following, FFT analysis was performed using the Analysis function ->Signal Processing -> FFT. The function outputs various tables and graphs, and we exported the power spectra (Frequency (Hz) vs. Power (dB)) as shown in Figure 2. The power (dB) measurements for each 118 FCS autocorrelation spectra obtained from the n=24 patient samples were extracted, as 1D vectors, and subjected to Scikit-learn ML classification algorithms with their respective binary patient labels (Healthy vs. Control). Further, the power spectral images as shown in **Figure 2 B and D**, for healthy and cancer patient samples, respectively, were obtained for the 118 spectra, and subjected to classification by various Deep Learning neural networks, as discussed below.

**Scikit-learn Statistical ML Classifiers:** Machine Learning analysis was performed using binary classifiers from the Scikit-learn python library (Pedregosa et al., 2011). Additionally, AdaBoost classifier was used as an ensemble learner to enhance the predictive performance of the RF classifier. The cross-validation was set to 10-fold (CV=10). All shown results used a 50:50 training: testing split for stringent classification conditions. Lower test splitting resulted in higher performance. ML classification was performed on all N= 118 complete FCS power spectra of n= 24 patients. All sample data and codes are provided in the GitHub link repository (See Data and Code Availability Section). The hyperparameters of the ML classifiers were tuned as follows: The following frequencies were identified as the most optimal frequencies at which the patients' FCS power spectral classification were best distinguished by all ML classifiers: F1 = 0 Hz, F2 = 0.237, F3 = 1.896 Hz, F4 = 2.60699 Hz, F5= 2.72549 Hz. These frequencies were identified by manual brute-force searching and by visual inference of the power spectra for regions where prominent power fluctuations were observed distinguishing the two patient groups.

**Support Vector Machines (SVM):** class sklearn.svm.SVC (C=1.0, break_ties=False, cache_size=200, class_weight=None, coef0=0.0, decision_function_shape='ovr', degree=3, gamma='scale', kernel='linear', max_iter=-1, probability=False, random_state=None, shrinking=True, tol=0.001, verbose=False)

**Random Forest (RF) Classifier:** Unlike the other ML classifiers discussed below, the performance was constrained to a selected set of optimal frequencies (as determined using SVM's optimal performance, shown below).
*RandomForestClassifier(max_depth=6, max_features='sqrt', min_samples_leaf=3, min_samples_split=10, n_estimators=50)*

**Multilayer Perceptron (MLP):** The hyperparameters for the MLP algorithm were tuned as follows:
*class sklearn.neural_network.MLPClassifier(hidden_layer_sizes=(30,30,30), activation='relu', *, solver='adam', alpha=0.01, batch_size='auto', learning_rate='constant', learning_rate_init=0.001, power_t=0.5, max_iter=200, shuffle=True, random_state=None, tol=0.0001, verbose=False, warm_start=False, momentum=0.9, nesterovs_momentum=True, early_stopping=False, validation_fraction=0.1, beta_1=0.9, beta_2=0.999, epsilon=1e-08, n_iter_no_change=10, max_fun=15000).*

**Convolutional Neural Networks:** The following are convolutional neural networks (Deep Learning algorithms) performed on the N=118 FCS power spectral images obtained from the patient samples, as explained above.

**Image CNN:** The power spectra images were reshaped to uniform sizes using IMAGE_SHAPE = (224, 224) and an 80:20 training: testing validation was used. Hyperparameter tuning was performed as follows: optimizer = tf.keras.optimizers.Adam(lr=1e-2) model.compile( optimizer=optimizer, loss='categorical_crossentropy', metrics=['acc']).
The training and cross-validation was set to epochs=50, verbose=1.
*model = tf.keras.Sequential([ hub.KerasLayer('https://tfhub.dev/google/tf2-preview/mobilenet_v2/feature_vector/4', output_shape=[1280], trainable=False), tf.keras.layers.Dropout(0.4), tf.keras.layers.Dense(train_generator.num_classes, activation='softmax') ]) model.build([None, 224, 224, 3])*

**Table 2**. Image CNN layers-architecture and model hyperparameters.

```
_________________________________________________________________
Layer (type)              Output Shape        Param #
=================================================================
keras_layer (KerasLayer)  (None, 1280)        2257984
_________________________________________________________________
dropout (Dropout)         (None, 1280)        0
_________________________________________________________________
dense (Dense)             (None, 2)           2562
=================================================================
Total params: 2,260,546
Trainable params: 2,562
Non-trainable params: 2,257,984
```

**ResNet Image Classification:** Resnet models 101, 18, and 34 were attempted and all yielded nearly identical performance results. The model number was insensitive to our results. The hyperparameters were set as follows: bs = 64 (batch size): if your GPU is running out of memory, set a smaller batch size, i.e., 16 sz = 224 (image size)**,** learn.fit_one_cycle (10, max_lr=slice(1e-3,1e-3))**,** 80% for training and 20% for validation**, and** learn = cnn_learner (data, models. resnet101, metrics=accuracy).

**Quantum Neural Network:** The study further involves benchmarking Quantum Convolutional Neural Network (QNN) algorithm (Sengupta and Srivastava, 2021) for comparing performance with classical convolutional neural networks (CNNs) used above. An 80:20 and 60:40 training: testing split were used as validation sizes on the N=118 spectral images, as shown in the confusion matrices in Figure 4C and 4D, respectively. The workflow for the QNN algorithm was as follows:
1. Input raw data using Keras
2. Filtering the dataset to only 3 s and 6 s
3. Downscales the images to fit in a quantum hardware.
4. Treating and removing contradictory examples
5. Convert binary images to Cirq circuits
6. Convert the Cirq circuits to a TensorFlow quantum circuit

**Pre-processing/downscaling** - OpenCV library was leveraged for morphological transformations , the method was employed majorly for handling noise and detection of intensity collisions. Further Image denoising (Buades et al., 2011) and scaling using Python-OpenCV library was implemented to the entire dataset for standardization. **Circuit design approach:** A two layered circuit(qubit=3) was designed for the datasets with hinge loss as loss function and ADAM (adaptive learning rate optimization) optimizer instead of stochastic gradient descent-based optimizer being computationally inexpensive and easy to implement.

**Table 3.** QNN Hyperparameters.

| Parameter(s) | Value |
|---|---|
| Layer | PQC |
| Output Shape | (None,1) |
| Param | 32 |
| Model | Sequential |
| Loss Function | Hinge |
| Optimizer | ADAM |
| Evaluation Metrics | Hinge Accuracy |
| **Epoch** | 10 |
| **Batch size** | 32 |
| # of samples | 118 |

**Linear and Nonlinear Feature Extraction:** The FCS power spectra data files were imported as a csv file containing the frequency as the first column, and subsequent columns corresponded to the Power (dB) measurements of each patient FCS power spectrum. The following feature selection methods were used to quantify additional spectral features which may be useful in downstream pattern analysis or prospective studies.

**Principal Component Analysis (PCA):** PCA was performed as a linear dimensionality reduction on the N=118 FCS power spectra using the Scikit-learn package in Google Colab (See GitHub link for code).

**Nonlinearity Dimensionality Reduction:** Diffusion Map and Isomap were used as nonlinear dimensionality reduction algorithms to observe whether any nonlinear features could help distinguish the healthy and cancer patients-derived power spectra. The Python codes for both algorithms are provided as Jupyter notebooks in the GitHub link.

**Multifractal Analysis:** Multifractal analysis was used as a spectral feature extraction method to assess fractal dynamics in the time-series data. The Holder exponent was calculated using OriginPro, by taking the log-log plot of the power spectra and using a linear fit analysis on the log-log plot to estimate the slope $\alpha$ (i.e., the Holder exponent). The Hurst exponent was also computed using the MATLAB Wavelet Transform Modulus Maxima (WTMM) package using the [dh1, h1, cp1, tauq1] = dwtleader (Power) function, where Power corresponds to the imported 1D vector (column) with the power (dB) measurements for a single power spectral sample. A table must be made with the computed Hurst exponent for all N=118 samples.

**RESULTS and DISCUSSION**

The results of our pilot study demonstrates that FCS coupled with AI-algorithms has the potential to become an accurate diagnostic cancer screening tool that can be integrated in liquid biopsies and clinical precision oncology. As shown in **Figure 2**, the autocorrelation FCS spectra of healthy patient EVs sample and cancer patient EVs sample is shown in **Figure 2A** and **2C**, respectively. The characteristic inverted S-like autocorrelation curve is observed in both cases. There are prominent fluctuations seen in the tail ends of the curve. However, we predicted that using the Fast-Fourier Transform (FFT) to obtain the power fluctuations of the autocorrelation function in frequency space would provide a more robust screening tool to distinguish healthy patients' spectra from cancer patients' spectra. As such, the corresponding FFT-Power spectra for healthy and cancer patient EVs samples are shown in **Figure 2B** and **2D**, respectively. A power decay is observed in the fluctuations over the frequency range. We predicted machine learning algorithms, including binary classifiers and spectral-image based convolutional neural networks (CNNs) may be capable of better spotting patterns and signatures distinguishing the two patient groups using the processed power spectra.

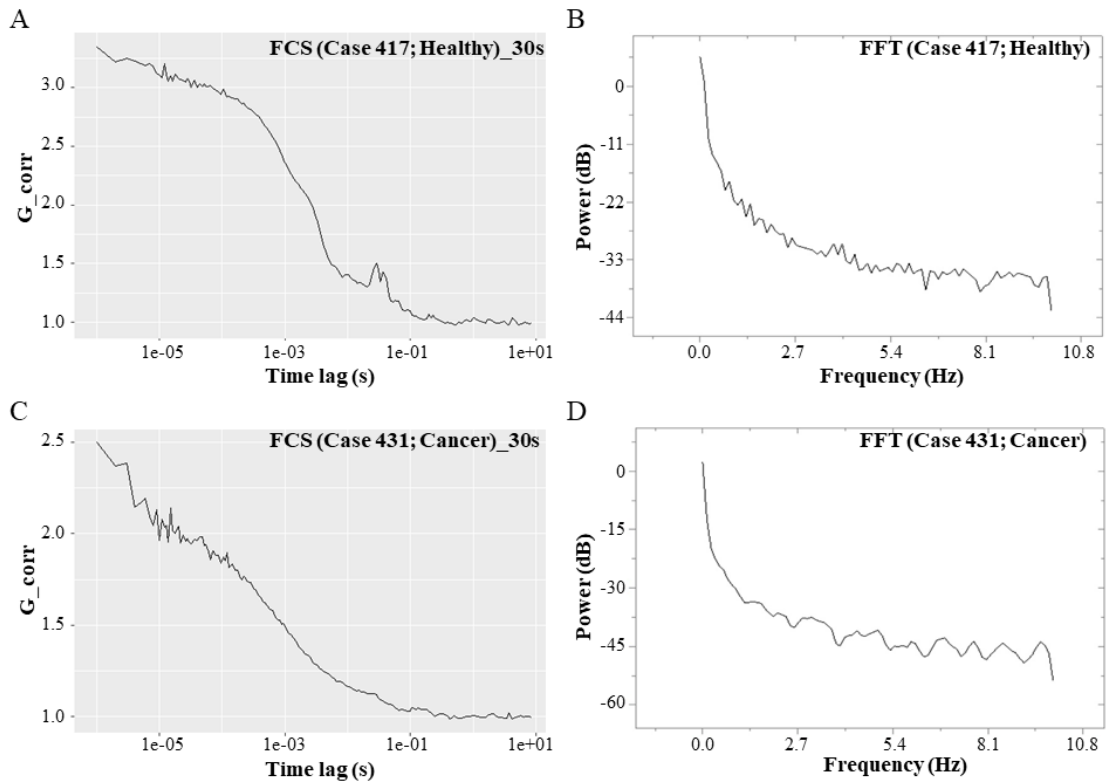

**Figure 2. FCS Autocorrelation spectra and processed Power Spectra.**
A) Autocorrelation spectrum of Healthy control, with 30 s acquisition time.
B) Power spectrum of Healthy control corresponding to Fig 2A.
C) Autocorrelation spectrum of Cancer patient sample with 30 s acquisition time.
D) Power spectrum of Cancer sample corresponding to Fig 2C.
Representative data are displayed from patient 417 (healthy) and patient 431 (cancer affected).

In **Figure 3**, we see the statistical performances of various Scikit-learn ML classifiers on the processed FCS power spectra. All confusion matrices shown in **Figure 3** were subjected to a 50:50 training: testing validation split, for more stringent testing conditions. Their performances were found to be of higher accuracy with lower training sizes. The testing was also constrained to the power (dB) values at five selected frequencies, found to be the most optimal set of values for the performance of the SVM and RF classifiers. In **Figure 3A**, the performance of a hyperparameter-tuned multilayer perceptron (MLP) neural network with 30 layers, is shown as a confusion matrix. The classification accuracy was found to be 0.73 (i.e., 73%) with a 10-fold cross-validation (CV) score of 61.33+/- 11.85%. The sensitivity was found to be 0.33 and the specificity was found to be 0.875, using the confusion matrix using the online confusion matrix calculator (See Data and Code Availability Section). The average f1-score, an additional measure of the ML's accuracy, was found to be 0.61 and 0.71, for the healthy and cancer groups, respectively.

In **Figure 3B** and **3C**, we see the cross-validation learning curve and confusion matrix for the ML performance of the AdaBoost Random Forest (RF) Classifier. The classification accuracy was found to be 0.9091, with a mean-square error of 0.09. The precision scores were found to be 0.92 and 0.91 for the healthy and cancer spectra, respectively, while the f1 scores were found to be 0.88 and 0.91, respectively. The sensitivity and specificity of the RF performance were 0.733 and 0.975, respectively. The RF classifier performed near the classification accuracy of our image-based CNNs when the complete power spectra were subjected to classification (i.e., near 80% accuracy) (data not shown). We found that its optimal performance is obtained by constraining the algorithm towards the five selected frequencies which optimized the SVM performance. Amidst all tested ML classifiers, the RF demonstrated the highest accuracy, sensitivity, and specificity. Further testing with larger patient-sample cohorts is required to validate its clinical potential.

In **Figure 3D**, the performance of the Support Vector Machine (SVM) classifier is shown. The classification accuracy was found to be 0.618 with a 10-fold cross-validation score of 69.33+/-7.42%. The average f1-scores were 0.55 and 0.60 for the healthy and cancer groups, respectively. The sensitivity and specificity were found to be 0.30 and 0.80, respectively. SVM showed the poorest classification performance in terms of accuracy, amidst the three ML binary classifiers, as further explained by the poor linear separability seen in PCA analysis (See **Figure 5F**). SVM uses a hyperplane to linearly separate data points from the two patient groups into two separate classes, and hence, we suggest that such linear separability is a poor metric for classifying such complex spectral data.

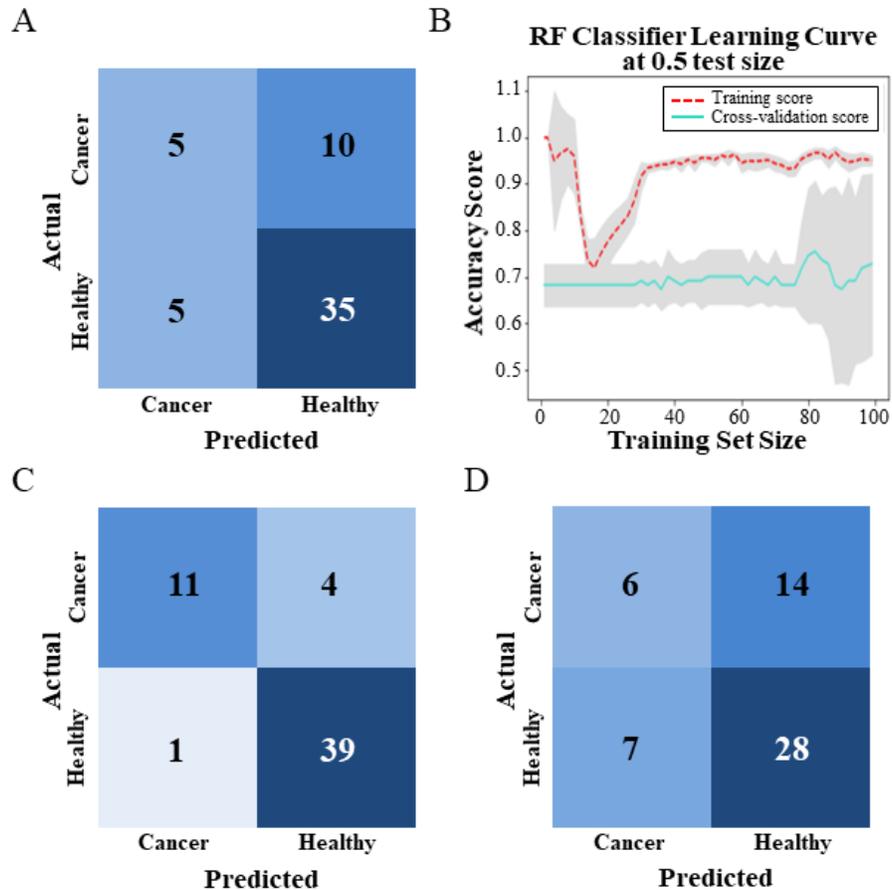

**Figure 3. ML classification on FCS Power spectra.** The validation test size was set to 50:50 training: testing split for all ML classifiers performance assessment. All tests were initially performed using an 80:20 split but for stringent conditions were subjected to a 50:0 split, wherein the performance in the 80:20 or 70:30 were better than that of 50:50. Selected power spectral frequencies (Hz) for analysis: 0, 0.237, 1.896, 2.60699, 2.72549.

A) Multilayer Perceptron: Average precision for control group was 0.64 and for cancer group was 0.70. The average recall was 0.60 and 0.73, while the f1-scores were 0.61 and 0.71, respectively. The classification accuracy was 0.73. The 10-fold CV score was 61.33+/- 11.85%.

B and C) Random Forest Classifier: Of a test set of 55 sample spectra out of 105 power spectra, Accuracy: 90.91 %, MSE: 0.0909, CV = 10-fold, cross-validation score was found to be $56.00 \pm 21.90\%$. Precision score for control (healthy) groups and cancer groups was 0.92 and 0.91, respectively, with an average recall of 0.85 and 0.91, respectively. The average f1-scores were 0.88 and 0.91, respectively.
D) Support vector machines (linear kernel): the classification accuracy 61.82%. Ten-fold CV score was found to be 69.33+/-7.42%. The average precision scores for the control and cancer groups were found to be 0.56 and 0.59, respectively. The average recall scores were 0.55 and 0.62 and the f1-scores were 0.55 and 0.60, respectively.

To further validate our findings from the ML binary classifiers, we exploited image-based AI algorithms, namely CNNs on the FCS power spectra images. As shown in **Figure 4A**, a Tensorflow image CNN's performance is shown in the confusion matrix. The classification accuracy was 0.826, with a 10-fold CV score of 0.74. The f1-score was found to be 0.875, whereas the sensitivity and specificity were 1.00 and 0.56, respectively. Although a perfect sensitivity is obtained, the accuracy and specificity are not as optimal and hence, such classification results should be interpreted with caution. **In Figure 4B**, the cross-validation and learning curve for the Image CNN in **Figure 4A** is shown. As seen, with increasing training steps, the validation curve (in orange) stabilizes to a near 0.74 CV accuracy score. **Figure 4C** and **4D** display the confusion matrices for the performance of a Quantum CNN adopted from Sengupta and Srivastava (2021), with a training: testing validation sizes of 80:20 and 60:40, respectively. In **Figure 4C**, the classification accuracy was found to be 0.833, while the f1-score was determined to be 0.882. The precision score was 0.938, while the sensitivity and specificity were both found to be 0.833, matching with the classification accuracy. In **Figure 4D**, with the 60:40 validation size, the classification accuracy and f1-score were obtained as 0.78 and 0.864m respectively. The precision score, sensitivity, and specificity were found to be 0.854, 0.875, and 0.400, respectively. As shown, while the QNN results seem to be of a reasonable classification performance with a lower validation size, when more stringent conditions are applied, there is a loss in accuracy and specificity. As such, the results remain inconclusive and require a larger patient cohort for clinical validation. Lastly, **Figure 4E** shows the learning curve for the ResNet 34 CNN, which obtained equivalent results as those obtained for the Image CNN in **Figure 4A** (i.e., the classification accuracy for the ResNet was 82.6%). Based on this preliminary evidence, we can conclude that the classical CNNs and the QNN perform very similarly on our dataset with near 80% classification accuracy. Given the complexity of the power spectra, we suggest these findings support the concept that CNNs be used as a cross-validation tool along with the RF classifier discussed above, in larger patient cohort screening in prospective studies.

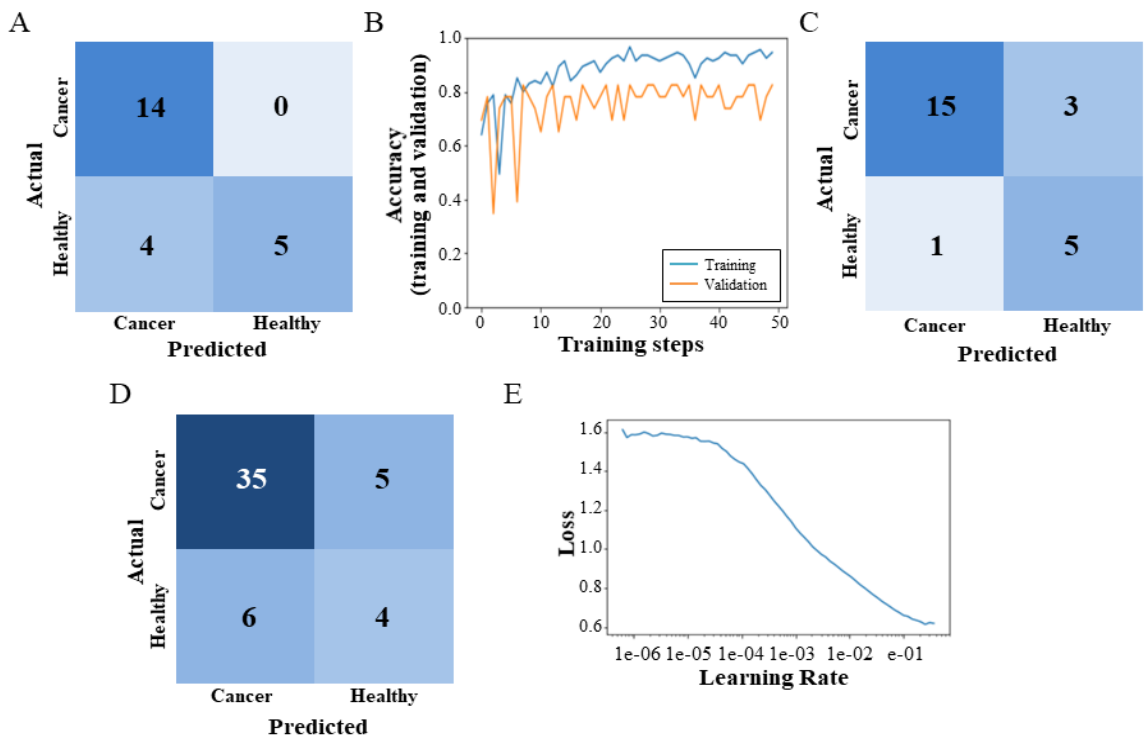

**Figure 4. Convolutional Neural Networks' Performance on Power Spectra Images (N=118).**

A) Image CNN confusion matrix. Final accuracy: **82.61%**. Final loss with 10-fold CV: 0.74.
B) 10-fold Cross-validation curve for Image CNN in A. Blue: Training performance. Orange: Validation.
C) Quantum Neural Network (QNN) performance on an 80:20 training: testing validation set. Classification Accuracy: **83.33%.**
D) QNN performance on a 60:40 validation set. Classification Accuracy: **78 %.**
E) Residual Neural Network Image classification (on Power spectra). Learning Curve for Resnet 34. Classification Accuracy: **82.6%.** The generic Image CNN in **Figure 4A** and the Resnet 34 both performed equally with a classification accuracy near 82%, confirming the method's consistency.

Finally, we explored some feature extraction algorithms to determine whether certain dimensionality reduction algorithms or multifractal characteristics of the complex FCS spectra can be used to distinguish the patient groups in prospective ML analyses. As shown in **Figure 5A**, Diffusion Map shows a clear separation between the two patient groups' power spectra. In contrast, **Figure 5B**, shows that Isomap, a local multi-dimensional scaling without the local optima, performs poorly in separating the two patient groups. **Figure 5C** and **5D**, display two types of multifractal analyses, the Holder exponents (log-log plot scaling determined by the linear best-fit/correlation for the power spectra) and the Hurst index computed using the wavelet-based WTMM algorithm, are also poor classifiers of the two patient groups. Thus, no unique multifractal feature could distinguish the patient spectra, as further supported by the poor classification performance of the Hurst index scores of the two patient groups for the N=118 spectra, using the AdaBoost RF classifier shown in **Figure 5E**. Lastly, **Figure 5F** shows that there is no linear separability in the power spectra by the linear dimensionality reduction algorithm PCA.

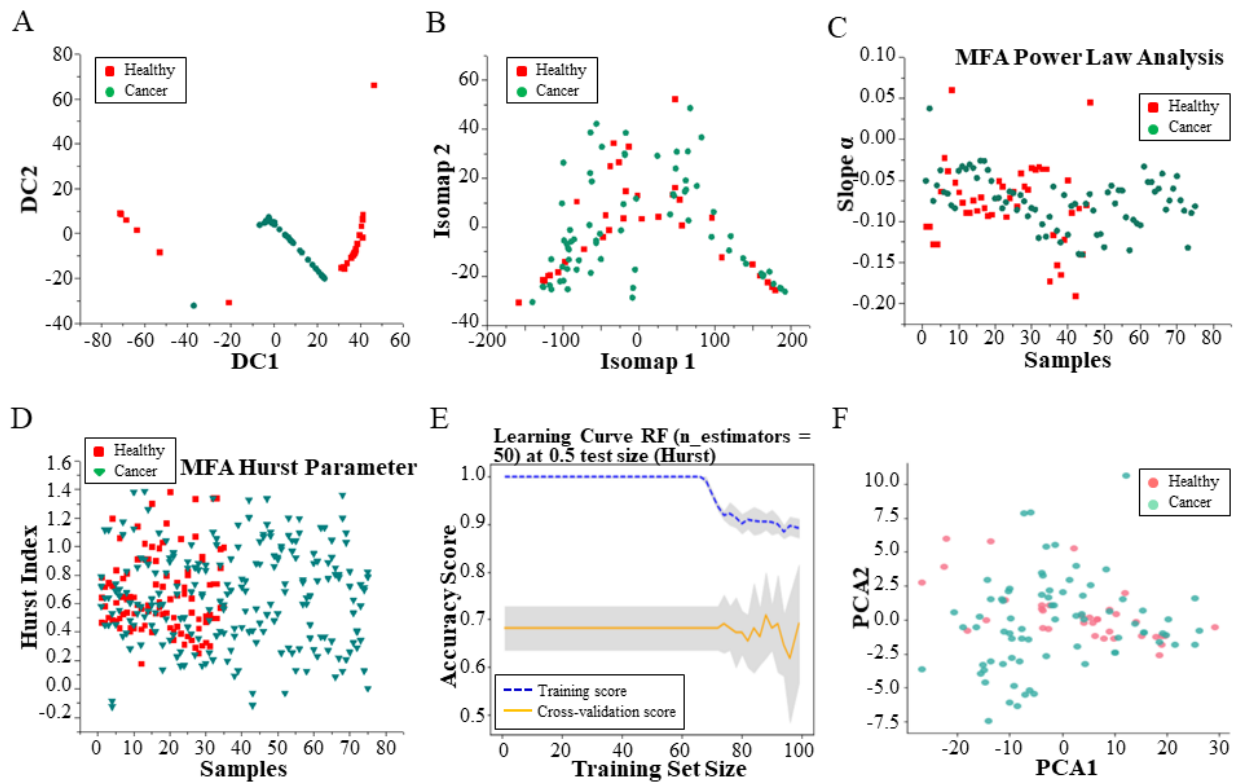

**Figure 5. Feature Extraction on FCS Power spectra.**
A) Diffusion Map, a type of nonlinearity reduction, performed on the FCS power spectra. A clear separation is seen by their first two Diffusion Components (DC).

B) Isomap nonlinear dimensionality reduction on the FCS power spectra.
C) Multifractal Power Law analysis on FCS power spectra, wherein the slope denotes the Holder exponent.
D) Multifractal Hurst exponent analysis on the power spectra.
E) Random Forest Learning curve with a 10-fold cross validation on the Hurst exponent data.
F) Linear dimensionality reduction by Principal Component Analysis (PCA) on the power spectra.

These preliminary tests of feature extractors show that Diffusion Map may hold potential in optimizing the image-based CNNs or ML classifiers in prospective studies due to their clearer separability of the two patient groups' power spectra. Further, there remains a vast amount of nonlinear feature extraction methods which were not tested in our pilot study, including but not limited to, graph spectral clustering algorithms, spectral algorithmic complexity estimates, Gaussian processes, nonlinear neighborhood component analysis, and multivariate information-theoretic measures. we strongly suggest the testing of these nonlinear feature extraction methods in prospective analyses with larger patient cohorts.

**LIMITATIONS**

As shown by the classification results and their cross-validation scores, we should always keep in mind the limitations of these ML tools. Further, there are limitations in the Deep Learning frameworks of the CNNs used as well. For instance, adding a little random noise to CNN can largely fool its image classification. Flipping an image that was not in training, can also overthrow the algorithm's classification to false discovery (i.e., false positives). Further, there remains the black box problem in Deep Learning, making the process of pattern detection ambiguous although useful. These algorithmic limitations suggest that the algorithms' performance is specific to the training datasets, as indicated by the 10-fold cross validation scores, and imply that they highly depend on the model-system of interest. To overcome these barriers, a larger patient sample size is fundamentally required to validate the clinical benefits and relevance of our study. Regardless, the results should be treated as that of an interdisciplinary pilot study pioneering the coupling of FCS spectra, AI, and EVs-based cancer screening with high accuracy and promising results as seen in the RF's performance. A plausible explanation for the 90% classification accuracy in the RF classifier and not higher performance could be that benign mass patients were categorized as healthy for the ML training and assessment. Thus, given the vast heterogeneity and complexity of the tumor samples we analyzed within our pilot study of n=24 patients, with their distinct cell of origin/tissue subtypes, we can safely agree that our results warrant further analysis given its high statistical performance metrics for some algorithms like the RF classifier and the CNNs. The quantum machine learning showed near equal accuracy with the CNN, and hence, we conclude there were no additional advantages provided by Quantum machine learning. Given that Quantum optimized hardware and resources are needed for such quantum machine learning, as far as CNN-like algorithms are concerned, our study suggests prospective studies with larger cohorts of patients for clinically-relevant assessments could simply adhere to the use of classical CNNs for validation of our pilot results.

Further, it should be noted that in contrast to the data-driven statistical machine learning algorithms we have utilized herein, there are various model-driven AI approaches better-suited for complex feature analyses and forecasting patterns from the temporal features of complex time-series datasets not investigated herein. Some examples of such algorithms include recurrent neural networks such as liquid neural networks and Hopfield neural networks. There are certain biochemical limitations which were screened for during the FCS measurements, such as the clustering or clumping of EVs. The presence of such large aggregates/clumps were screened by the emergence of large spikes in the fluctuation intensity spectra (i.e., FCS counts) during the measurements. They could be additionally filtered manually by selecting time-windows omitting their presence, which was not needed in our case due to the careful analysis by the ABIF technician. Future studies should also investigate time-resolved spectroscopies with label-free (unstained) EVs.

**PROSPECTIVE STUDIES AND APPLICATIONS:** As mentioned, in our previous study, we have already explored vibrational spectroscopies such as Raman and FT-IR. Herein, we explored for the first-time the use of FCS, a time-resolved spectroscopy technique, to quantify, characterize and distinguish cancer patient-derived EVs from healthy patients-derived EVs. Future studies should further expand on our findings with larger cohorts of patients including cancers of distinct tissue subtypes and stages/grades. Further, there remains many other spectroscopic methods which can be coupled to patient-derived EVs and AI, including mass spectrometry techniques, surface-enhanced Raman spectroscopy (SERS), Terahertz spectroscopy, and high-energy spectroscopies, to name a few. There may be other types of time-resolved spectroscopy, or the above-mentioned spectroscopies can be adapted to time-resolved methods (e.g., Raman time-lapse imaging).

As mentioned, in contrast to the data-driven methods exploited herein, causal inference models like RNNs, namely, liquid cybernetics (i.e., liquid neural networks), LSTM, or reservoir computing, and Hopfield neural networks, and neuro-symbolic computation methods should be exploited in future large-scale time-series analyses when dealing with more than hundreds of patients (Maass et al., 2002; Verstraeten et al., 2007). These initiatives could be useful in the automated scientific/pattern discovery of complex patients-derived EVs spectra and the molecular fingerprinting of the time-series EVs power fluctuations in the future of personalized cancer nanomedicine. We have strictly focused our analyses on statistical ML-based classification. Further analyses should extend to physics and model-driven AI approaches for causal discovery, prediction, and forecasting, as discussed above. Prospective studies should also explore the applicability of Quantum Random Forest classifiers or Quantum Decision Trees (an ensemble of which becomes the RF) on the power spectra. Domain-free and model-independent feature selection algorithms optimized for minimal loss of algorithmic complexity should also be employed in prospective pattern analyses.

To conclude, our experiments are part of different pilot studies performed in the field of early cancer detection and interdisciplinary classification of patient-derived EVs. In the specific this pilot study warrants further advancement of the presented pairing of time-resolved spectroscopic techniques and artificial intelligence in the characterization of cancer patients-derived EVs. The presented approach may help in disease prevention and therapy management by serving as a candidate for non-invasive, diagnostic, and prognostic blood-based clinical screening. Our findings suggest such applied intelligence may bear fruits in the progression of computational systems oncology and diagnostic precision medicine.

**DATA AND CODE AVAILABILITY**
All codes and sample datasets obtained in this experiment are made available in the GitHub link below.

GitHub link: https://github.com/Abicumaran/FCS_EVClassification

Multifractal analysis: WTMM toolbox guidelines in MATLAB to extract Hurst scaling exponent:
https://www.mathworks.com/help/wavelet/ug/multifractal-analysis.html
https://onlineconfusionmatrix.com/ (to calculate the sensitivity and specificity from the confusion matrices)

**ACKNOWLEDGEMENTS.**
We are grateful to Ayat Salman for her assistance with the Ethical Committee approvals. Fluorescence correlation spectroscopy measurements were carried out by Joel Ryan at the McGill Advanced BioImaging Facility (ABIF, RRID: SCR_017697).

**STATEMENT OF ETHICS**
Patients were recruited in accordance with an approved ethics protocols by the Ethics Committee of the McGill University Health Centre (MP-37-2018-3916 and SDR-10-057). Patients signed consents were obtained before enrolment in the study.

**CONFLICT OF INTEREST STATEMENT**
The authors declare no conflict of interest.

**FUNDING SOURCES**
This work was financially supported by Giuseppe Monticciolo and the Morris & Bella Fainman Family Foundation. The funders had no role in study design, data collection and analysis, decision to publish, or preparation of the manuscript.

**APPENDIX.**

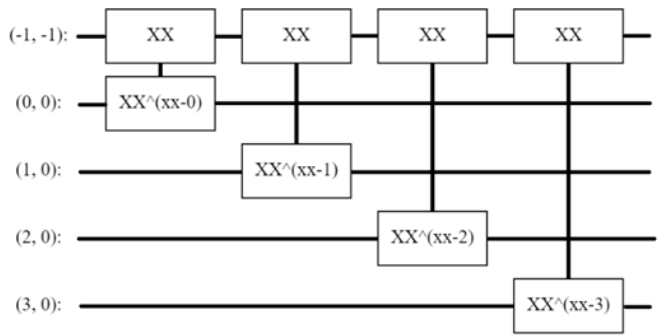

**QUANTUM ML CIRCUIT.** The circuit from the training samples in the first iteration of the 2-layer circuit, reproduced from Sengupta and Srivastava (2021). For further details of the QNN, refer to the citation.